\documentclass[aps,prl,twocolumn,groupedaddress,showpacs,floatfix]{revtex4-1}
\usepackage{graphicx}
\usepackage{amsmath}
\usepackage{textcomp}
\usepackage{mathrsfs}
\usepackage{amsfonts}

\begin{document}

\newcommand{\beq}{\begin{equation}}
\newcommand{\eeq}{\end{equation}}
\newcommand{\barr}{\begin{eqnarray}}
\newcommand{\earr}{\end{eqnarray}}
\newcommand{\bseq}{\begin{subequations}}
\newcommand{\eseq}{\end{subequations}}
\newcommand{\vett}[1]{\mathbf{#1}}
\newcommand{\uvett}[1]{\hat{\vett{#1}}}
\newcommand{\ket}[1]{|#1\rangle}
\newcommand{\mat}[4]{\left[
\begin{array}{cc}
#1 & #2 \\ #3 & #4 \\ 
\end{array}
\right]}

\title{Demonstration of local teleportation using classical entanglement}

\author{Diego Guzman-Silva$^1$}
\author{Robert Br\"uning$^2$}
\author{Felix Zimmermann$^1$}
\author{Christian Vetter$^1$}
\author{Markus Gr\"afe$^1$}
\author{Matthias Heinrich$^1$}
\author{Stefan Nolte$^1$}
\author{Michael Duparr\'e$^2$}
\author{Andrea Aiello$^{3,4}$}
\author{Marco Ornigotti$^1$}
\author{Alexander Szameit$^1$}

\affiliation{$^1$Institute of Applied Physics, Friedrich-Schiller Universit\"at Jena, Max-Wien Platz 1, 07743 Jena, Germany}
\affiliation{$^2$ Institute of Applied Optics, Abbe Center of Photonics, Friedrich-Schiller-Universit\"at Jena, Max-Wien-Platz 1, 07743 Jena, Germany}
\affiliation{$^3$ Max Planck Institute for the Science of Light, G\"unther-Scharowsky-Strasse 1/Bau24, 91058 Erlangen, Germany}
\affiliation{$^4$ Institute for Optics, Information and Photonics, University of Erlangen-Nuernberg, Staudtstrasse 7/B2, 91058 Erlangen, Germany}

\email{marco.ornigotti@uni-jena.de}

\date{\today}

\begin{abstract}
Teleportation is the most widely discussed application of the basic principles of quantum mechanics. Fundamentally, this process describes the transmission of information, which involves transport of neither matter nor energy. The implicit assumption, however, is that this scheme is of inherently nonlocal nature, and therefore exclusive to quantum systems. Here, we show that the concept can be readily generalized beyond the quantum realm. We present an optical implementation of the teleportation protocol solely based on classical entanglement between spatial and modal degrees of freedom, entirely independent of nonlocality. Our findings could enable novel methods for distributing information between different transmission channels and may provide the means to leverage the advantages of both quantum and classical systems to create a robust hybrid communication infrastructure.
 \end{abstract}

\maketitle

\section{Introduction}
Quantum teleportation allows a sender to share information with a receiver by encoding it into a quantum state that both parties can access. In general, the counter-intuitive aspects of this type of protocols stem from the fact that, in contrast to all conventional forms of communication, the state itself is not being transmitted between the two \cite{ref1}. Following the first landmark experiments employing photonic quantum states as information carriers \cite{ref2,ref3a,ref3}, teleportation has been implemented in a wide variety of different physical systems \cite{ref4,ref5,ref6,ref6bis}. Teleportation is therefore considered to be a fundamental building block for quantum communication and computation architectures. At the core of its ability to distribute information through potentially vast and far-reaching networks is entanglement, which allows the nonlocal interaction that is necessary for a transfer of information to occur \cite{ref7}.

For many years, entanglement was viewed as the epitome of a quantum effect, seemingly lacking any analogue in classical systems. Only very recently it became clear that the underlying algebraic structure giving rise to entanglement can indeed be reproduced in the classical domain \cite{ref8,ref9,ref10,ref11,ref12,ref13,ref14}. Very recently, moreover, state transfer protocols based on the classical entanglement concept have also been proposed \cite{refBoyd}. As it turns out, the key distinction between classical and quantum systems is not their capability to support entanglement, but rather the (non-) locality of the same \cite{ref12}. As subsequent works have demonstrated \cite{ref15,ref16,ref17,ref18}, these notions may help to infuse classical physics with the potentialities offered by quantum entanglement to drive novel and interesting applications. 
In this work, we extend the concept of teleportation beyond its well-known quantum context. To this end, we propose and experimentally demonstrate a photonic setting in which classical entanglement can be harnessed to teleport information between the spatial and modal degrees of freedom of a purely classical light field. While nonlocality remains the differentiating feature between quantum and classical configurations, our results illustrate that teleportation-based protocols can bridge the gap between these two realms.

\section{Theoretical background}
On a conceptual level, teleportation involves at least three degrees of freedom, which henceforth will be labelled as A, B and C. Whereas an entangled Bell state is established between A and B, the information one desires to transmit is contained in C. The actual teleportation process is facilitated by the application of a controlled-NOT (C-NOT) operation with respect to A and C, followed by a Hadamard operation and the subsequent projection onto a Bell state [7]. For the case of quantum teleportation, these degrees of freedom are represented by qubits, while in the classical case they are called ÒcebitsÓ \cite{ref14}. By adopting DiracÕs notation, the input state can be written in the following form (see Supplementary Material)

\beq
\ket{\Psi}=\frac{1}{\sqrt{2}}\left[\left(\alpha\ket{0}_C+\beta\ket{1}_C\right)\otimes\ket{\Phi^+}_{AB}\right],
\eeq

where the cebit C resides in a two-dimensional Hilbert space. The complex coefficients $\alpha$ and $\beta$ carrying the information can be normalized such that $|\alpha|^2+|\beta|^2=1$. Here, $\ket{\Phi^+}_{AB}$ represents a classically entangled Bell state. The aforementioned combination of C-NOT and Hadamard gates then yields the following output state:

\barr\label{eq2}
\ket{\Psi_{out}}&=&\frac{1}{2}\Big[\ket{00}_{CA}\otimes\left(\alpha\ket{0}_B+\beta\ket{1}_B\right)\nonumber\\
&+&\ket{01}_{CA}\otimes\left(\alpha\ket{1}_B+\beta\ket{0}_B\right)\nonumber\\
&+&\ket{10}_{CA}\otimes\left(\alpha\ket{0}_B-\beta\ket{1}_B\right)\nonumber\\
&+&\ket{11}_{CA}\otimes\left(\alpha\ket{1}_B-\beta\ket{0}_B\right)\Big].
\earr

The entanglement, initially shared between the cebits A and B, has been transferred to C and A. To realize the teleportation of information between the cebits C and B, $\ket{\Psi_{out}}$  is projected onto one of the four basis states spanning the joint Hilbert space of the two entangled cebits A and C. A readout of the information contained in the cebit B will therefore give us access to the information initially encoded in the cebit C.  It is worth noting that Eq. \eqref{eq2} is completely independent from the character of the degrees of freedom involved. Rather, it is the choice of entanglement that contextualizes the associated process in the classical, local, world, or in the nonlocal realm of quantum mechanics. This observation leads us to interpret Eq. \eqref{eq2} as the teleportation equation. It directly follows that the gate sequence of the teleportation scheme is universally valid, irrespectively on the nature of the entanglement employed.

\section{Experiment}

To implement such a protocol, we employ the experimental setup shown in Fig. \ref{figure1}, where the information can be teleported from the path (C) to the spatial (B) Hilbert space. Light from a 594 nm He-Ne laser is sent to a custom- designed wave plate (RPWP) to synthetize a single classically entangled beam with radial polarization, corresponding to $\ket{\Phi^+}_{AB}$ \cite{ref10}. Having prepared the cebits A and B for our classical implementation of teleportation, the third cebit C can be readily established between two identical copies of such classically entangled beam, propagating along different paths. One path represents $\ket{0}_C$, and the other $\ket{1}_C$, respectively \cite{ref12}.  Experimentally, this is realized by means of a 50/50 beam splitter (BS1), and two neutral-density filters ($F_{\alpha}$ and $F_{\beta}$) placed at each output channel of BS1 encode the information in the cebit C. 

The output state $\ket{\Psi_{out}}$ described by Eq. \eqref{eq2} therefore corresponds to the two output channels of BS2. Note that each of its four unique mode structures can be individually extracted with a simple measurement operation. Consider, for example, the first term in Eq. \eqref{eq2}, which can be written, in terms of optical modes, as follows (see Supplementary Material):
\begin{figure}[!t]
\begin{center}
\includegraphics[width=0.5\textwidth]{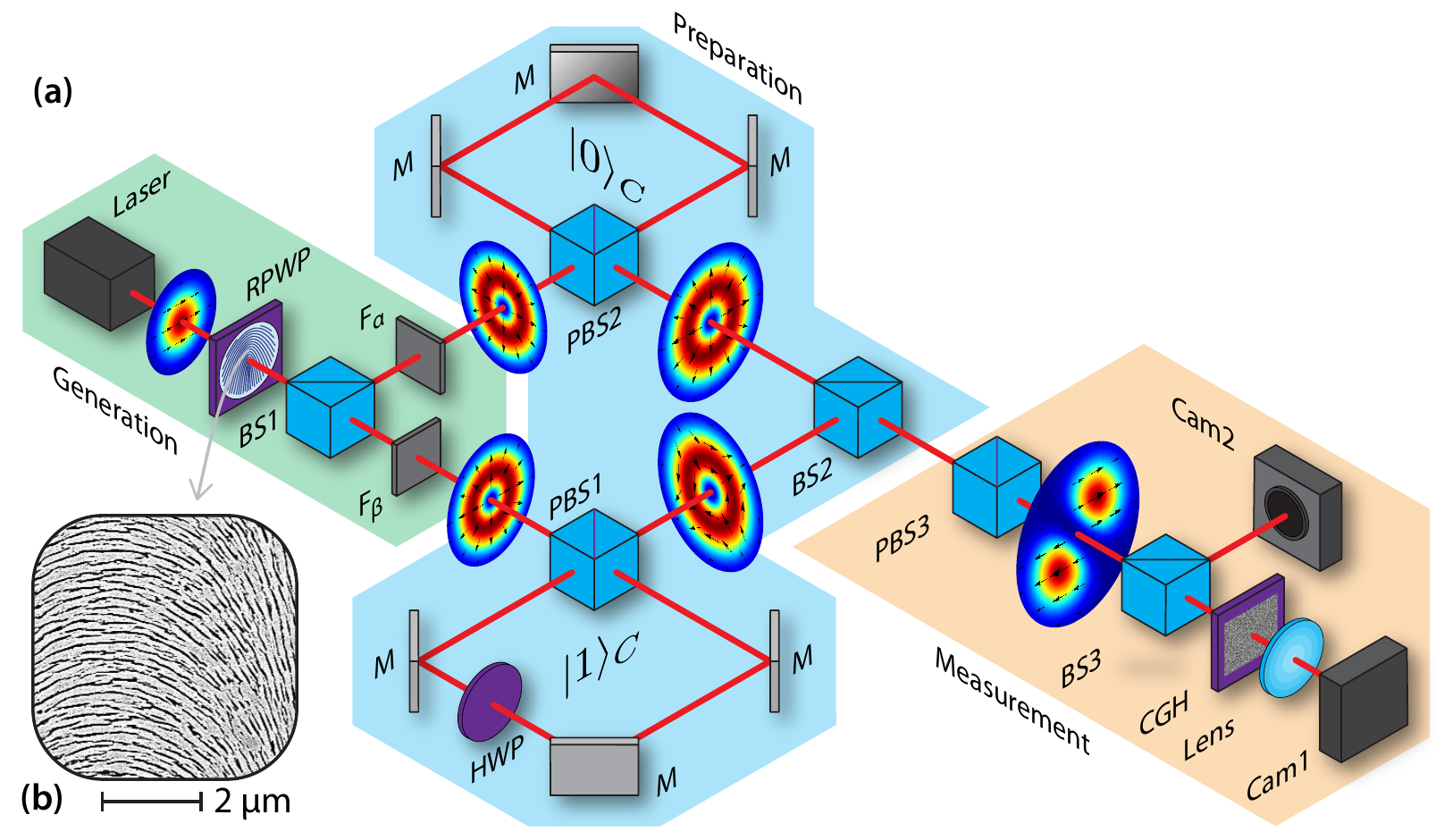}
\caption{(a) Generation: A continuous-wave He-Ne laser is prepared in a classically entangled state with radial polarization by a rotating polarization wave plate (RPWP). The initial state $\ket{\Psi}$ is then obtained using a beam splitter (BS1). Two density filters ($F_{\alpha}$ and $F_{\beta}$) are used to encode the information in the cebit C. Preparation: The initial state $\ket{\Psi}$ is then sent through a C-NOT gate, realized using two Sagnac interferometers with a polarizing beam splitter. The lower interferometer, corresponding to  $\ket{1}_C$ also contains a half-wave plate (HWP) to rotate the polarization. The two parts of the beam are then recombined through a second beam splitter (BS2), which implements a Hadamard operation for the cebit C. Measurement: The correct output state is selected by choosing the x-polarization (PBS3) of the lower output channel of BS2. The reflected beam from a third beam splitter (BS3) is sent to a CCD camera (Cam2) for direct acquisition of the intensity profile and angle measurement. The transmitted beam from BS3 is instead sent to the modal decomposition stage, consisting of a computer generated hologram (CGH) a lens and a second CCD camera (Cam1). (b) Exemplary scanning electron microscope image of a nanograting-based polarization rotating wave plate such as the RPWP used to generate the classically entangled beam with radial polarization.}
\label{figure1}
\end{center}
\end{figure}
\barr
\ket{00}_{CA}&\otimes&\left(\alpha\ket{0}_B+\beta\ket{1}_B\right)\nonumber\\
&=&\uvett{x}[\alpha\psi_{10}^u(x,y)+\beta\psi_{01}^u(x,y)].
\earr

Therein, $\psi_{10}$ and $\psi_{01}$ are the first-order Hermite-Gaussian solutions of the paraxial equation \cite{ref20}, and $\uvett{x}$,$\uvett{y}$ denote the horizontal and vertical polarization eigenstates, respectively. The path cebit C has been implemented by labelling the upper and lower beam with the superscript $u$ and $l$, respectively. Then,$\ket{0}_C$ corresponds to the beam being in the upper path, while $\ket{1}_C$ corresponds to the beam being being in the lower path \cite{ref12}.  A closer inspection to the above equation reveals that in order to project the output state $\ket{\Psi_{out}}$ onto the state $\ket{00}_{CA}$, it is sufficient to take the $\uvett{x}$-polarization eigenstate of the beam in the upper path, i.e., $\ket{0}_C$. This guarantees access to the information encoded in the cebit B. In our experimental setup, this projection operation is realized by placing a polarizer (PBS3) at the output of BS2 corresponding to $\ket{0}_C$.
\begin{figure}[!t]
\begin{center}
\includegraphics[width=0.5\textwidth]{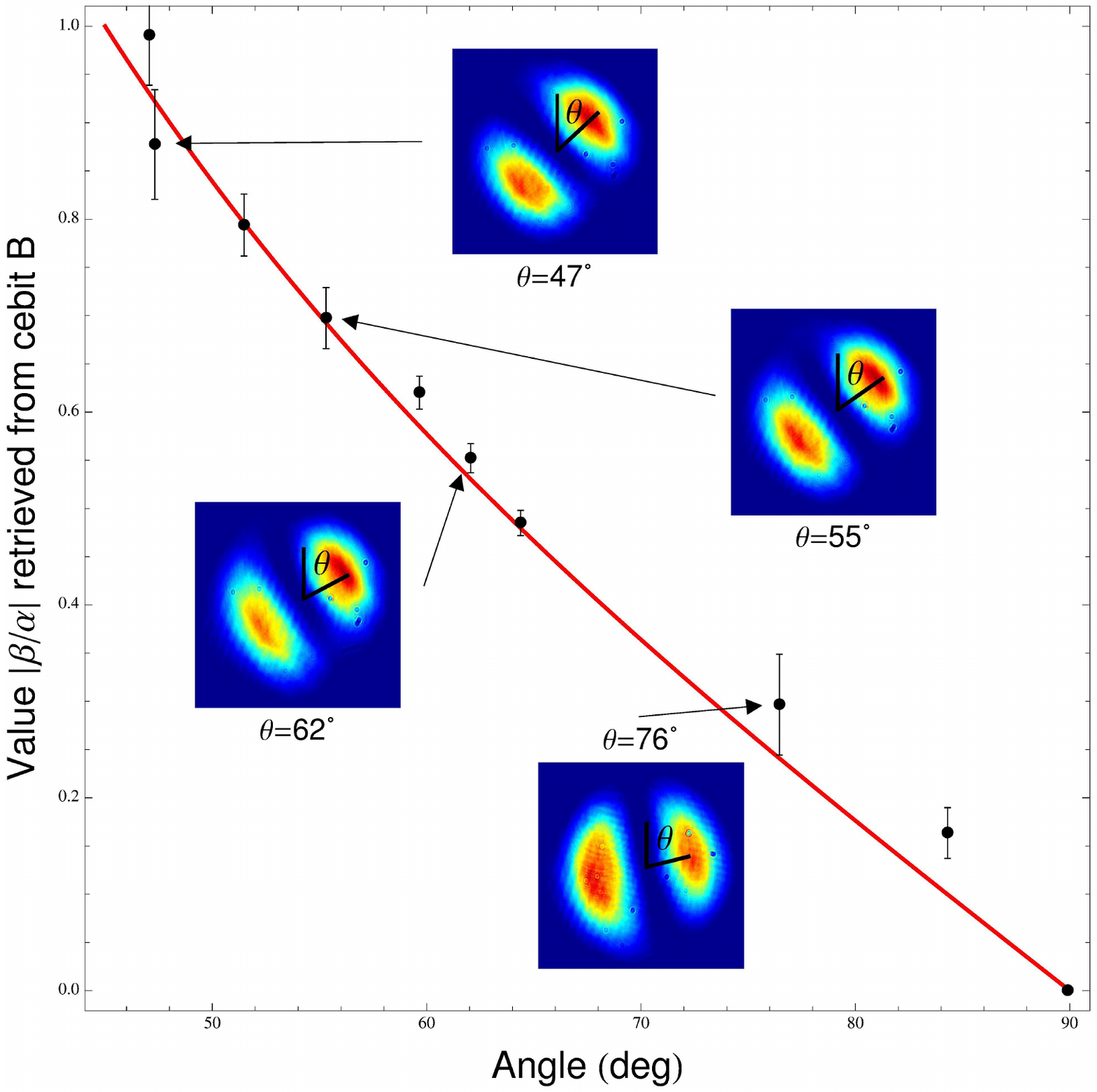}
\caption{Estimation of the ratio $\beta/\alpha$ initially encoded in the cebit C as retrieved from cebit B by measuring the angle formed by the intensity distribution acquired from the CCD camera Cam2 with the vertical axis. As can be seen, the retrieved angles (black points) are in good agreement with the theoretical prediction (red solid line), namely $\cot\theta=|\betaÚ\alpha|$. The insets show the acquired intensity distribution (single realization) corresponding to$\theta = 47^{\circ}, 55^{\circ}, 62^{\circ}$ and $76^{\circ}$, respectively. }
\label{figure2}
\end{center}
\end{figure}
In terms of optical beams, the projected state $_{CA}\langle 00|\Psi_{out}\rangle$ assumes the following form:
\beq\label{eq4}
M(x,y)=\alpha\psi_{10}(x,y-y_0)+\beta\psi_{01}(x,y-y_0).
\eeq

The beam $M(x,y)$ is then split into two parts by a beam splitter (BS3). While the reflected beam is directly acquired by a CCD camera (Cam2), the transmitted beam is instead sent through a computer-generated hologram (CGH) that decomposes it in its different Hermite-Gaussian constituents \cite{ref21}. Subsequent imaging onto a second CCD camera (Cam1) allows for the extraction of both the amplitude and the phase of the modal coefficients $\alpha$ and $\beta$. 

\begin{figure}[!t]
\begin{center}
\includegraphics[width=0.5\textwidth]{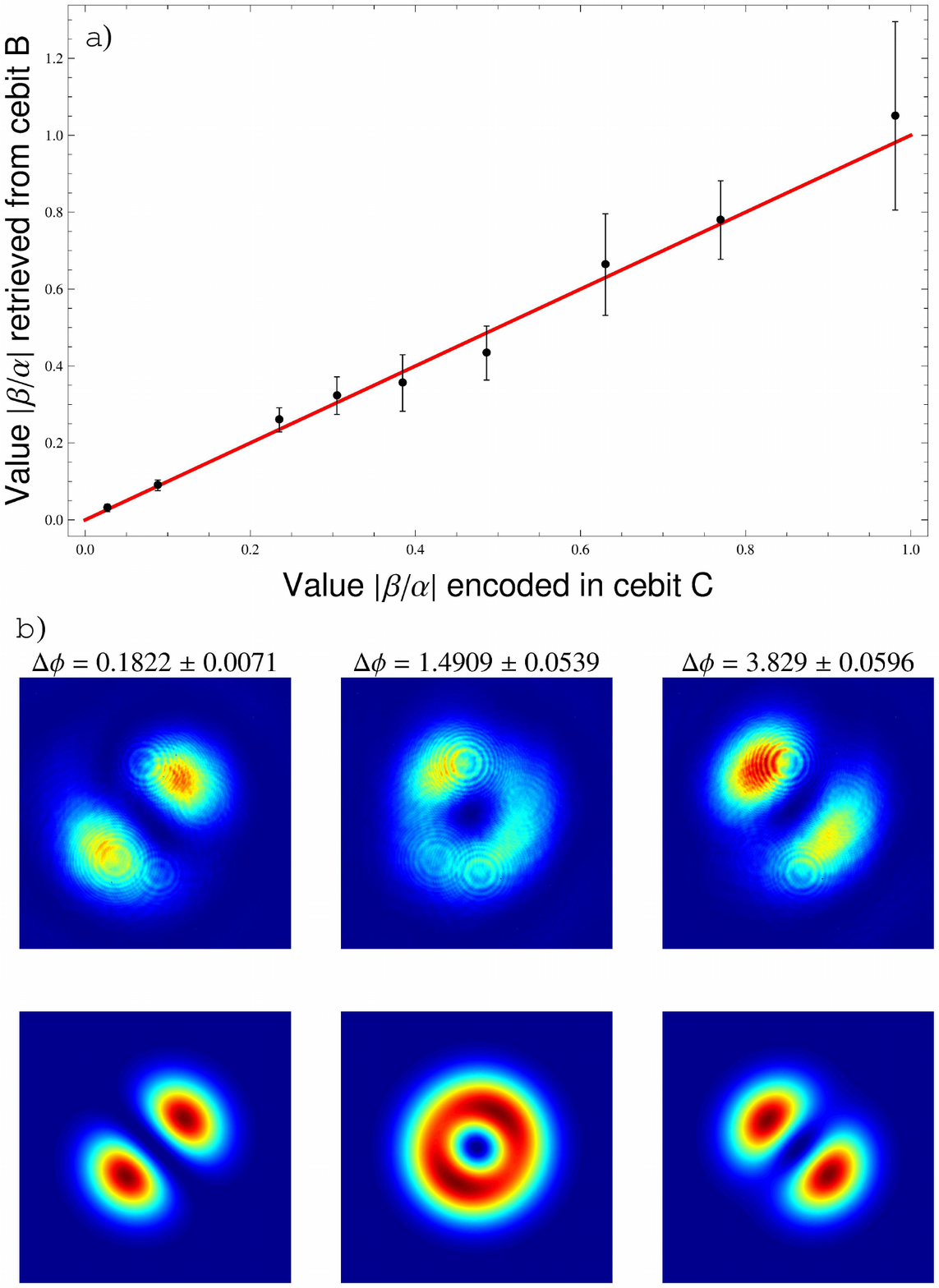}
\caption{(a) Plot of the ratio $|\beta/\alpha|$ as retrieved from cebit B as a function of the ratio $|\beta/\alpha|$initially encoded into cebit C. (b) Comparison of the retrieved intensity distributions (upper row) and the theoretical predictions (lower row) corresponding to different values of the phase difference $\Delta\phi$ encoded into $\alpha$ and $\beta$ as retrieved from cebit B. To encode the phase information into the cebit C, the filter $F_{\alpha}$ has been tilted by $2.5^{\circ}, 5^{\circ}$ and $10^{\circ}$, respectively, while the filter $F_{\beta}$ has been left unchanged.  The correspondent retrieved phases are displayed above the acquired intensity distributions, together with the corresponding error intervals.}
\label{figure3}
\end{center}
\end{figure}

The information encoded in the spatial cebit B can be then retrieved in two different ways. If the phase difference between $\alpha$ and $\beta$ is negligible, the intensity distribution acquired by Cam2 is a Hermite-Gaussian mode $\psi_{01}$ rotated by an angle $\theta$ with respect to the $y$-axis. The value of the modal coefficients can be then directly obtained from the acquired image by measuring the rotation angle $\theta$ with a digital algorithm (see Methods). 
Exemplary results of this measurement, together with the acquired intensity distributions corresponding to $\theta = 47^{\circ}, 55^{\circ}, 62^{\circ}$ and $76^{\circ}$, are shown in Fig. \ref{figure2} and follow the theoretical prediction $|\betaÚ\alpha|=\cot\theta$. If, on the other hand, the phase difference between $\alpha$ and $\beta$ is not negligible, and a full modal decomposition is required to extract the information, as illustrated in Fig. \ref{figure3}. 

Clearly, the choice of the spatial degree of freedom as the target cebit for teleportation allows for an easy and reliable measurement and a classical communication network, thus opening the possibility for new ways to reroute information between initially disjoint Hilbert spaces.

\subsection{Generation of the classically entangled state}

To generate a classically entangled beam with radial polarization, we employed a rotating polarization wave plate (RPWP), like the one shown in Fig. \ref{figure1}(b). This special optical component was custom-fabricated using a femtosecond laser writing technique and it consists of a $\lambdaÚ2$ plate with locally varying anisotropy. The local anisotropy is realized by inscribing periodic sub-wavelength nano-gratings in the bulk of a fused silica substrate \cite{ref22,ref23}. By tuning the polarization of the inscribing laser, the orientation of the nanograting can be set as desired. This offers a great degree of control over the polarization pattern imprinted on the impinging beam on the micron scale. 

\subsection{Estimation of the modal coefficients via angle measurements}

When the phase difference between $\alpha$ and $\beta$ is negligible, the value of these modal coefficients can be retrieved by directly acquiring the modal function $M(x,y)$.  According to Eq. \eqref{eq4}, in fact, the intensity distribution acquired by Cam2 consists of a Hermite-Gaussian mode $\psi_{01}$ rotated by an angle $\theta$ with respect to the vertical axis. To measure such an angle, we first evaluate numerically, from the retrieved image, the line $L$ joining the center of mass of each lobe. Then, we compute the slope of the line perpendicular to $L$, to directly obtain the angle formed by the intensity distribution $|M(x,y)|^2$ with the vertical axis. The values of $\alpha$ and $\beta$ can be therefore retrieved by using the relation $\cot?\theta=|\betaÚ\alpha|$ (see Supplementary Material). In order to provide an accurate measurement of the angle $\theta$, the results presented in Fig. \ref{figure2} have been averaged over fifteen acquired intensity distributions corresponding to the same set of parameters $\alpha$ and $\beta$. The error bars in Fig. \ref{figure2}, moreover, correspond to the standard deviation calculated with such a set of experimental data.

\subsection{Estimation of the modal coefficients by modal decomposition}
When the phase difference between $\alpha$ and $\beta$ is not negligible, a full modal decomposition technique is needed to retrieve the information encoded in the spatial cebit B. With this technique, the modal function $M(x,y)$ described by Eq. \eqref{eq4} is decomposed into the Hermite-Gaussian modes $\psi_{10}$ and $\psi_{01}$. To do that, we employ the correlation filter method (CFM) technique (see Supplementary Material) \cite{ref21}, which basically implements the scalar product between the modal function described by Eq. \eqref{eq4} and the Hermite-Gaussian basis function $\psi_{lm}$, i.e.,

\beq\label{eq5}
\langle\psi_{lm}|M\rangle=\int\int_{-\infty}^{\infty}\psi^*_{lm}(x,y)M(x,y)dxdy.
\eeq

Therefore, since the Hermite-Gaussian functions $\psi_{10}$ and $\psi_{01}$ are the complete basis set that spans the two dimensional Hilbert space associated to cebit B, this method allows to extract the information encoded in this cebit by simply calculating $\langle\psi_{01}|M\rangle=\alpha$ and $\langle\psi_{10}|M\rangle=\beta$, respectively. This is realized in our experimental setup by using a computer-generated hologram (CGH) with appropriately designed transmission functions \cite{ref21}. The integration appearing in Eq. \eqref{eq5} has been then carried out using a $2f$ system, which performs the Fourier transform of the incoming signal. The  intensity distribution in the focal plane therefore provides access to the amplitude of the coefficient $\alpha$ and $\beta$. With this method, it is also possible to retrieve the relative phase difference $\Delta\phi$ between $\alpha$ and $\beta$ by means of a three-way interference between different replicas of the two mode functions with known phase difference. In addition, in order to provide real-time access to both $\alpha$ and $\beta$ simultaneously, we employed an angular multiplexing technique similar to the one described in Ref. \cite{ref21}.

Although the results presented in Fig. \ref{figure3} are in very good agreement with the expected ones, it has to be noted that the measurement error increases with the measured ratio $\beta/\alpha$. To obtain a good correlation signal, in fact, the scalar product \eqref{eq5} needs to be evaluated at a single point (e.g., along the optical axis of the system), where the signal originating from the two Hermite-Gaussian modes are truly independent of one another. Fluctuations of the position of this point due to small fluctuations of the position where the two beams recombine (BS2) are the main source of measurement errors and can be minimized by ensuring a sufficient contrast in signal strengths between the modes.

\section{Conclusions}
With this work we have extended the concept of teleportation to the classical realm by implementing a photonic setting in which classical entanglement allows to teleport information between path and spatial degrees of freedom of a purely classical light field. In doing so, we have shown that teleportation is a general concept, that transcends the distinction between classical or quantum systems, and that non-locality ultimately differentiates between these two realms. Along these lines, our results may pave the way towards the realization of a hybrid classical-quantum communication infrastructure.

This work was supported by the Marie Curie Actions within the Seventh Framework Programme for Research of the European Commission, under the Initial Training Network PICQUE, Grant No. 608062. The authors further gratefully acknowledge financial support from the German Ministry of Education and Research (Center for Innovation Competence program, Grant No. 03Z1HN31) and the Deutsche Forschungsgemeinschaft (Grant No. NO462/6-1). M. H. was supported by the German National Academy of Sciences Leopoldina (grant LPRD 2014-03).


\begin{thebibliography}{99}


\bibitem{ref1} C. H. Bennett, G. Brassard, C. Cr\'epeau, R. Jozsa, A. Peres, and W. K. Wootters, ``\emph{Teleporting an unknown quantum state via dual classical and Einstein-Podolsky-Rosen channels}", Phys. Rev. Lett. \textbf{70}, 1895 (1993). 
\bibitem{ref2}   D. Bouwmeester, J.-W. Pan, K. Mattle, M. Eibl, H. Weinfurter and A. Zeilinger, ``\emph{A experimental quantum teleportation}", Nature \textbf{390}, 575 (1997).  
\bibitem{ref3a}L. Vaidman, ``\emph{Teleportation of quantum states}", Phys. Rev. A \textbf{49}, 1473 (1994).
\bibitem{ref3}	A. Furusawa, J. L. Sorensen, S. L. Braunstein, C. A. Fuchs, H. J. Kimble, E. S. Polzik, ``\emph{Unconditional quantum teleportation}", Science \textbf{282}, 706 (1998). 
\bibitem{ref4} M. Riebe, H. H\"affner, C. F. Roos, W. HŠnsel, J. Benhelm, G. P. T. Lancaster, T. W. K\"orber, C. Becher, F. Schmidt-Kaler, D. F. V. James and R. Blatt, ``\emph{Deterministic quantum teleportation with atoms}",   Nature \textbf{429}, 734 (2004).
\bibitem{ref5} H. Krauter, D. Salart, C. A. Muschik, J. M. Petersen, Heng Shen, T. Fernholz and E. S. Polzik, ``\emph{Deterministic quantum teleportation between distant atomic objects}",  Nature Phys. \textbf{9}, 400 (2013).
\bibitem{ref6} W. Pfaff, B. Hensen, H. Bernien, S. B. van Dam, M. S. Blok, T. H. Taminiau, M. J. Tiggelman, R. N. Schouten, M. Markham, D. J. Twitchen, R. Hanson, ``\emph{Unconditional quantum teleportation between distant solid-state quantum bits}",  Science \textbf{345}, 532 (2014).
\bibitem{ref6bis}A. Z. Khoury and P. Milman, \emph{Quantum teleportation in the spin-orbit variables of photon pairs}, Phys. Rev. A \textbf{83}, 060301(R) (2011).
\bibitem{ref7}  M. A. Nielsen and I. L. Chuang, \emph{Quantum Computation and Quantum Information} (Cambridge University Press, 2000).
\bibitem{ref8} R. J. C. Spreeuw, ``\emph{A classical analogy of entanglement}",  Found. Phys. \textbf{28}, 361 (1998).
\bibitem{ref9}  A. Luis,  ``\emph{Coherence, polarization, and entanglement for classical light fields}", Opt. Commun. \textbf{282}, 3665 (2009).
\bibitem{ref10} A. Holleczek, A. Aiello, C. Gabriel, C. Marquardt, and G. Leuchs, ``\emph{Classical and quantum properties of cylindrically polarized states of light}",  Opt. Express \textbf{19}, 9714 (2011).
\bibitem{ref11}  X.-F. Qian and J. H. Eberly, ``\emph{Entanglement and classical polarization states}",  Opt. Lett. \textbf{20}, 4110 (2011).
\bibitem{ref12} A. Aiello, F. T\"oppel, C. Marquardt, E. Giacobino, G. Leuchs, ``\emph{Quantum?like nonseparable structures in optical beams}",  New J. Phys. \textbf{17}, 043024 (2015).
\bibitem{ref13} P. Ghose and A. Mukherjee, ``\emph{Entanglement in classical optics}",  Rev. Theor. Sci \textbf{2}, 1 (2014).
\bibitem{ref14} R. J. C. Spreeuw, ``\emph{Classical wave-optics analogy of quantum-information processing}",  Phys. Rev. A \textbf{63}, 062302 (2001).
\bibitem{refBoyd} S. M. H. Rafsanjani, M. Mirhosseini, O. S. Magana-Loaiza and R. W. Boyd, \emph{State transfer based on classical noneparability}, Phys. Rev. A \textbf{92}, 023827 (2015).
\bibitem{ref15} F. T\"oppel, A. Aiello, C. Marquardt, E. Giacobino and G. Leuchs, ``\emph{Classical entanglement in polarization metrology}",  New. J. Phys. \textbf{16}, 073019 (2014).
\bibitem{ref16} C. V. S. Borges, M. Hor-Meyll, J. A. O. Huguenin, and A. Z. Khoury, ``\emph{Bell-like inequality for the spin-orbit separability of a laser beam}", Phys. Rev. A \textbf{82}, 033833 (2010).
\bibitem{ref17} E. Karimi, J. Leach, S.Slussarenko, B. Piccirillo, L. Marrucci, L. Chen, W. She, S. Franke-Arnold, M. J. Padgett, and E. Santamato, ``\emph{Spinorbit hybrid entanglement of photons and quantum contextuality}", Phys. Rev. A \textbf{82}, 022115 (2010).
\bibitem{ref18} L. J. Pereira, A. Z. Khoury, and K. Dechoum, ``\emph{Quantum and classical separability of spin-orbit laser modes}", Phys. Rev. A \textbf{90}, 053842 (2014).
\bibitem{ref20} O. Svelto, \emph{Principles of Lasers} (Springer, 2010).
\bibitem{ref21} T. Kaiser, D. Flamm, S.Schr\"oter, and M. Duparr\'e,  ``\emph{Complete modal decomposition for optical fibers using CGH-based correlation filters}", Opt. Express \textbf{17}, 9347(2009).
\bibitem{ref22} L. P. R. Ramirez, M. Heinrich, S. Richter, F. Dreisow, R. Keil, A. V. Korovin, U. Peschel, S. Nolte and A.T\"unnermann, ``\emph{Tuning the structural properties of femtosecond-laser-induced nanogratings}",  Appl. Phys. A \textbf{100}, 1 (2010).
\bibitem{ref23} M. Beresna, M. Gecevicius, P. G. Kazansky and T. Gertus, ``\emph{Radially polarized optical vortex converter created by femtosecond laser nanostructuring of glass}",  Appl. Phys. Lett. \textbf{98}, 201101 (2011).

%
\end{thebibliography}
\end{document}